\pdfoutput=1

\documentclass{article}
\usepackage{spconf,amsmath,graphicx}

\usepackage{amsfonts}
\usepackage{float}
\usepackage{cite}  

\def\x{{\mathbf x}}

\def\z{{\mathbf z}}
\def\u{{\mathbf u}}

\title{Synthesizer Preset Interpolation using Transformer Auto-Encoders}
\name{Gwendal Le Vaillant $^{1, 2}$, Thierry Dutoit $^{1}$}
\address{$^{1}$ Information, Signal and Artificial Intelligence (ISIA), University of Mons, Belgium \\
	$^{2}$ HE2B-ISIB Research Institute, Brussels, Belgium
}
\begin{document}
%
\maketitle
\begin{abstract}

Sound synthesizers are widespread in modern music production but they increasingly require expert skills to be mastered. 
This work focuses on interpolation between presets, i.e., sets of values of all sound synthesis parameters, to enable the intuitive creation of new sounds from existing ones.

We introduce a bimodal auto-encoder neural network, which simultaneously processes presets using multi-head attention blocks, and audio using convolutions.
This model has been tested on a popular frequency modulation synthesizer with more than one hundred parameters.
Experiments have compared the model to related architectures and methods, and have demonstrated that it performs smoother interpolations.
After training, the proposed model can be integrated into commercial synthesizers for live interpolation or sound design tasks.

\end{abstract}

\section{Introduction}

Sound synthesizers can generate audio signals whose timbre ranges from acoustic instruments to entirely novel sound textures.
They are ubiquitous in modern music production, and their use even defines some new music genres.
Synthesis processes are controlled using sets of parameters, called presets, which are usually large~\cite{auto_synth_prog_IEEE2018, esling2020flowsynth, sound2synth_2022}. They require expert knowledge to be created and handled, so that a lot of presets are provided by synthesizer manufacturers and developers themselves.

Our research addresses the problem of preset interpolation, 
for musicians to be able to discover new sounds in-between two reference presets, or to create smooth transitions from a preset to another. 
This requires to manipulate dozens of parameters simultaneously, with intricate relationships between parameters and synthesized audio, and interactions between parameters themselves.

While several recent works have used neural networks to match synthesizer presets with input sounds~\cite{auto_synth_prog_IEEE2018,  esling2020flowsynth, sound2synth_2022, presetgenVAE_2021, serumRNN_2021, inversynth2019}, 
ours is the first to formally focus on preset interpolation.
Its main contribution is a model that enables smoother interpolations, compared to related generative architectures.
It is also the first model that successfully handles synthesizer presets as sequences using Transformer~\cite{transformer_NeurIPS17} encoders and decoders, 
and models numerical synthesis parameters using Discretized Logistic Mixture (DLM) distributions~\cite{pixelCNNpp_iclr17}.

\section{Related Work}

\subsection{Neural Audio Synthesis}

Models such as WaveNet~\cite{wavenet2017nsynth} can be trained to synthesize raw audio waveforms using Convolutional Neural Networks (CNNs).
In order to prevent per-sample computations, recent works have tried to learn synthesis processes akin to commercially available music synthesizers.
Some are based on source-filter models~\cite{DDSP, vapar_synth_ICASSP2020}, whereas others~\cite{DDX7_ISMIR2022} model a differentiable FM (Frequency Modulation) synthesis architecture similar to a synthesizer named DX7.

However, these neural networks include the synthesis itself, and they are trained using gradient descent. 
Therefore, they can't be applied to existing commercial synthesizers~\cite{sound2synth_2022, presetgenVAE_2021, serumRNN_2021}, which rely on non-differentiable processes.

\subsection{Sound Matching}

Various neural network architectures have been successfully used to search for synthesis parameters that correspond best to an input sound.
Long Short-Term Memory (LSTM,~\cite{lstm_hochreiter1997}) neural networks have been used to infer a preset from Mel-Frequency Cepstral Coefficients (MFCCs)~\cite{auto_synth_prog_IEEE2018}.
Other architectures were based on CNNs to process audio spectrograms or raw waveforms~\cite{esling2020flowsynth, presetgenVAE_2021, inversynth2019}, or a combination of CNN, LSTM and Multilayer Perceptron (MLP) blocks to process different types of input audio features~\cite{sound2synth_2022}. 

Several works~\cite{auto_synth_prog_IEEE2018, presetgenVAE_2021, sound2synth_2022} have focused on the sound matching task for a software implementation of the well-established DX7 FM synthesis architecture.
It is known to be able to synthesize a wide variety of digital- and natural-sounding instruments~\cite{DDX7_ISMIR2022}, while being notoriously hard to handle considering the large amount of synthesis parameters (155). 
Experiments presented in this paper focus on this non-differentiable FM synthesis architecture.

\subsection{Generative Models}

Among the previously cited works, a few~\cite{esling2020flowsynth, presetgenVAE_2021} are based on generative models, which first encode input audio data $\x$ into a latent vector $\z$, then try to reconstruct the audio and infer the preset from $\mathbf{z}$. After training, latent vectors can be sampled from a prior distribution $p\left( \z \right) $ in order to generate new audio samples and new presets.

A common framework to learn both latent representations and a generative model is the Variational Auto-Encoder (VAE)~\cite{kingma2014autoencoding}.
It learns an approximate posterior distribution $q\left( \z|\x \right)$, which represents how $\x$ is encoded into the latent space, and a decoder model $p\left(\x, \z\right) = p\left(\x | \z\right) p\left(\z\right)$.
The encoded distribution $q\left( \z|\x \right)$ is usually Gaussian with a diagonal covariance matrix, 
i.e. $q\left( \z|\x \right) = \mathcal{N}(\z ; \mathbf{\mu}, \mathbf{\sigma}^2)$ 
where $\mathbf{\mu}$ and  $\mathbf{\sigma}^2$ are the outputs of an encoder neural network.
The latent prior  $p\left( \z \right)$ is usually set to $\mathcal{N}(\z ; 0, I)$,
while  $p\left(\x | \z\right)$ can be any distribution whose parameters are the outputs of a decoder neural network.
The loss $\mathcal{L}(\x)$ is an upper bound on the negative log-likelihood of the true data distribution:

\begin{equation}
	\mathcal{L} (\x) =
	\beta D_{KL}\left[q(\z|\x) \Vert p(\z) \right]
	- \mathbb{E}_{\z \sim q(\z|\x)} \left[ \log p(\x|\z) \right]
	\label{eq:vanilla_vae_loss}
\end{equation}

where $D_{KL}$ denotes the Kullback-Leibler divergence and $\beta$ controls the tradeoff between latent regularization and reconstruction accuracy~\cite{beta_VAE}.

The first term from (\ref{eq:vanilla_vae_loss}) can be considered as a regularization term, because it forces $q(\z|\x)$ to remain close to a multivariate standard normal distribution.
This prevents $\x$ inputs from being encoded as distributions with disjoint supports, such that the latent space should be continuous~\cite{kingma2014autoencoding} i.e. similar inputs should correspond to similar encoded distributions.
Moreover, if $\x$ has a much higher dimensionality than $\z$, then the latter can be considered as a compressed and meaningful representation of $\x$.

\subsection{Interpolation using Auto-Encoders}

Auto-encoder models, e.g. VAEs~\cite{beta_VAE} or adversarial auto-encoders~\cite{ae_interp_ACAI_2019}, can be trained to improve the interpolation between data points. 
However, such works are based on learned generators, in contrast to ours which focuses on learning how to interpolate presets for an external sound generator.
Some previous works~\cite{esling2020flowsynth, presetgenVAE_2021} about VAE-based sound matching have stated that they could perform interpolations, as any regularized VAE model does. However, the interpolation itself was not studied, and no quantified results were presented.

\section{Preset Interpolation}

\subsection{Synthesizer and Datasets}

Focusing on DX7 FM synthesis, we used a published database of approximately 30k presets~\cite{presetgenVAE_2021}, 
and randomly split it into a 80\%/10\% training/validation set and a 10\% held-out test set. 
The main volume, transpose and filter controls, which are not part of the FM synthesis process, were set to their default values, and left untouched.
Each preset then consists of 144 parameters, including \textit{Algorithm} which controls the discrete routing of signals between oscillators.
\textit{Algorithm} alone can completely change the synthesized timbre, such that it is arguably the most important FM synthesis parameter. 
However, it introduces highly non-linear relationships between presets and output sounds, so the most recent works~\cite{sound2synth_2022, DDX7_ISMIR2022} trained a different model for each \textit{Algorithm} value.
In our work, a single model was trained and the dataset includes the \textit{Algorithm} parameter.

\begin{figure}[hb!]
	\includegraphics[width=\columnwidth]{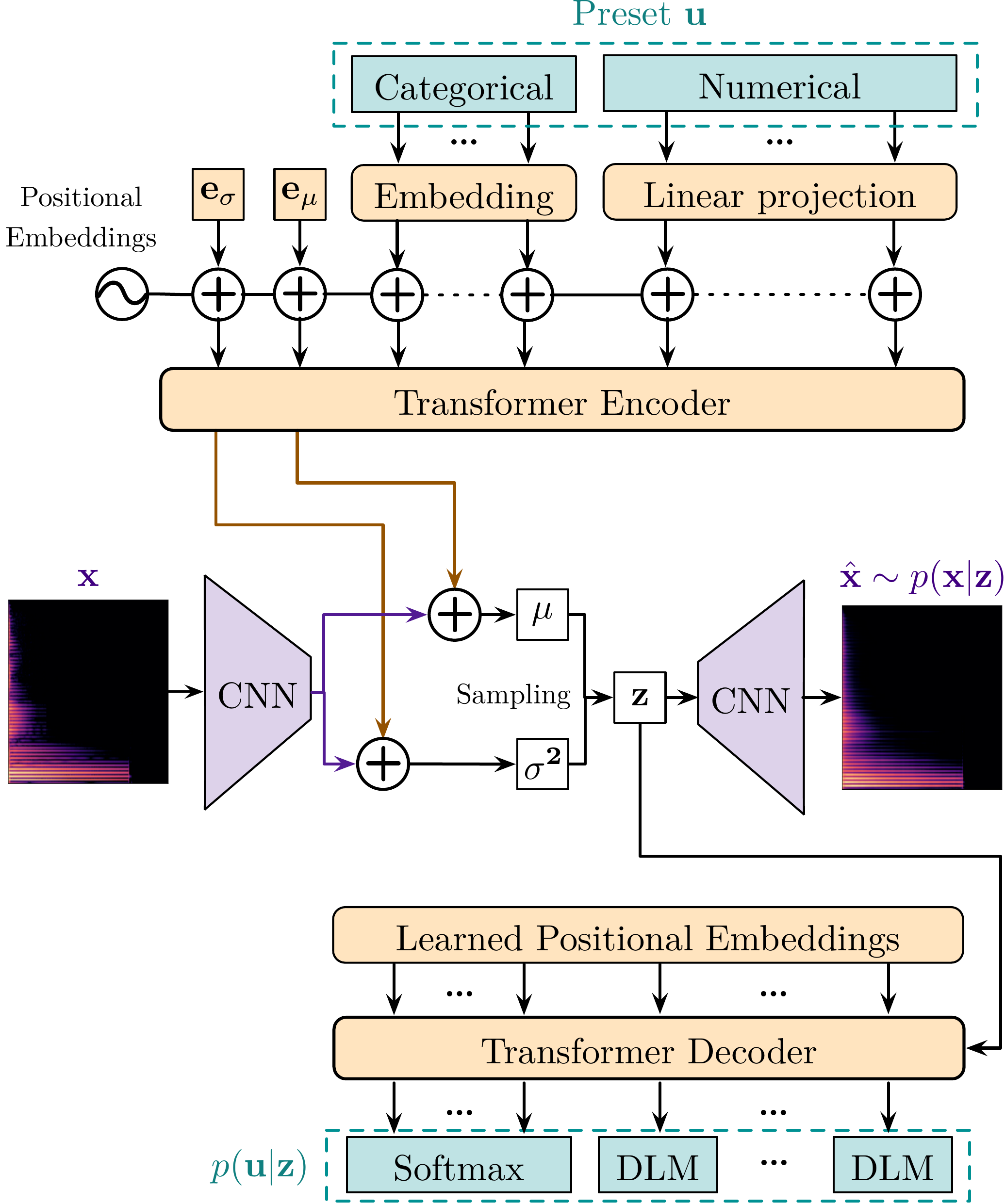}
	\caption{
		Overview of the SPINVAE (Synthesizer Preset INterpolation VAE) architecture.
	}
	\label{fig:general_arch}
\end{figure}

\begin{figure*}[!ht]
	\includegraphics[width=\textwidth]{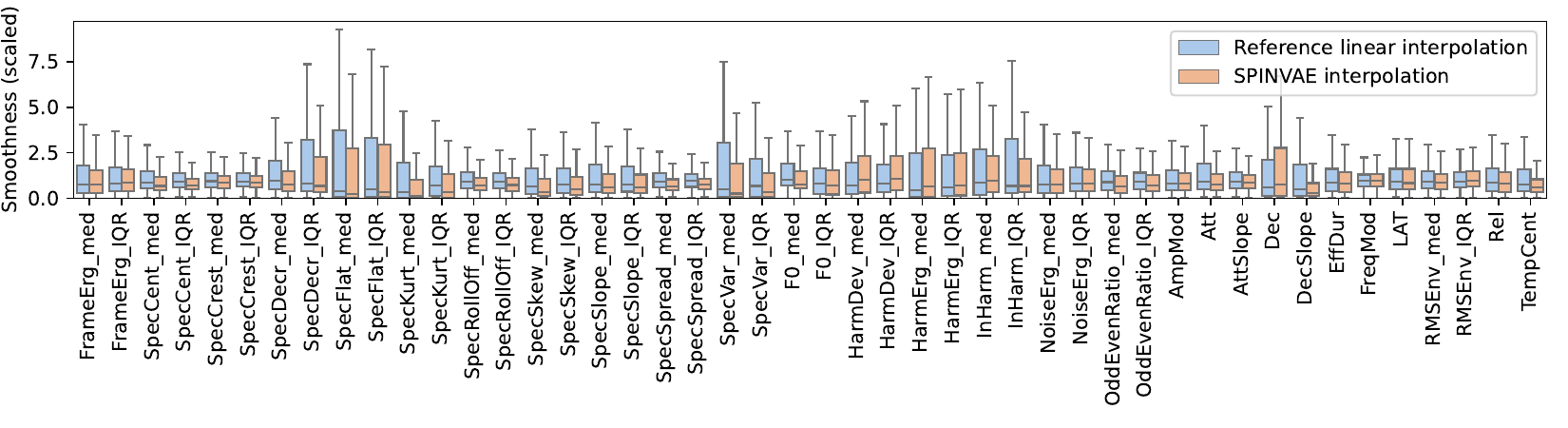}
	\caption{ 
		Interpolation smoothness (lower is better) for each audio feature extracted by Timbre Toolbox~\cite{timbretoolbox_jasa11}.
		For features to lie on a similar scale, values have been divided by the mean value of each feature for the reference model.
	}
	\label{fig:main_detailed_interp_results}
\end{figure*}

\subsection{Model}

Thanks to its latent space properties, the VAE framework is well suited to interpolation tasks.
For instance, a VAE could be trained to encode presets $\u$ into latent codes $\z$, and to decode them.
Nonetheless, it is not certain that the latent space would be continuous in the perceptual audio domain, 
i.e. that similar $\z^{(n)}, \z^{(m)}$ latent vectors would be decoded to $\u^{(n)}, \u^{(m)}$ that sound similar to a human.
Ideally, any $\z$ should hold meaningful compressed audio information.
Therefore, we employed a VAE that encodes and decodes presets $\u$ and spectrograms $\x$ (synthesized using $\u$) simultaneously.
The VAE loss becomes:

\begin{equation}
\begin{split}
	\mathcal{L} (\x, \u) = & \beta D_{KL}\left[q(\z|\x, \u) \Vert p(\z) \right] \\
						   & - \mathbb{E}_{q(\z|\x, \u)} \left[ \log p(\x, \u|\z) \right]
\end{split}
\label{eq:combined_vae_loss}
\end{equation}

We model $p(\x|\z)$ and $p(\u|\z)$ as independent distributions (Fig.~\ref{fig:general_arch}). 
The audio decoder $p(\x|\z)$ models each spectrogram pixel as a unit-variance Gaussian distribution.
The audio encoder and decoder are nine-layer CNNs with residual connections. 

The preset decoder $p(\u|\z)$ uses appropriate distributions for different synthesis parameters. 
Categorical parameters (e.g. \textit{Algorithm}, waveform type, etc.) are modeled by applying a softmax function on each output token.
Numerical parameters (e.g. frequency, attack time, etc.) are nonetheless discrete.
Therefore, we'll optimize the log-likelihood of DLM distributions~\cite{pixelCNNpp_iclr17}, 
which were originally proven effective to model pixel values. 
Such distributions are well suited to discrete numerical data with a limited range, because they compute probabilities using discrete bins, and tend to assign more probability to the lowest and highest bins.
Considering the histogram of numerical parameters values, three mixture components seemed appropriate.
Moreover, we extended the original DLM implementation to model parameters with different sets of discrete values (e.g. 8, 15, 100 quantized steps).

Among related works, only one~\cite{auto_synth_prog_IEEE2018} modeled presets as sequences, using LSTMs.
This was a natural choice because parameter values are highly dependent on others (e.g. \textit{Algorithm}).
However, our early tests using LSTMs had demonstrated poor performance and unstable training.
Therefore, presets are encoded and decoded using multi-head attention (Transformer, ~\cite{transformer_NeurIPS17}) blocks without masking, i.e. each hidden token can attend to tokens at any position in the sequence. 
Inspired by~\cite{transformer_vae_3Dmotion_ICCV21}, learnable input tokens $\mathbf{e_{\mu}}$ and $\mathbf{e_{\sigma}}$ are concatenated to the preset embeddings' sequence.
These two extra tokens are processed by the Transformer encoder, and the corresponding outputs are added to the CNN outputs.
On the decoder side, $\z$ is used to compute keys and values, while some learned input embeddings are used to compute queries.

The Transformer encoder and decoder are made of six layers each, and the latent dimension and Transformer hidden size have been empirically set to 256.
Implementation and training details are available in our source code repository\footnote{https://github.com/gwendal-lv/spinvae}.

\subsection{Audio Interpolation Metrics}\label{ssec:audio_interp_metrics}

After training and validation, the model has been used to compute 1.5k interpolation sequences between pairs of consecutive samples from the shuffled held-out test dataset (3k samples). 
First, two samples $(\x^{(n)}, \u^{(n)})$ and $(\x^{(m)}, \u^{(m)})$ 
are encoded into $\z^{(n)}=\mu^{(n)}, \z^{(m)}=\mu^{(m)}$. 
Then, a latent linear interpolation is performed to obtain $\left\lbrace \z_t, t \in [1, T] \right\rbrace $ vectors,
with $\z_{1} = \z^{(n)}$ and $\z_{T} = \z^{(m)}$.
Each $\z_t$ is finally decoded into a $\mathbf{u}_t$ preset, programmed into the synthesizer and rendered to audio. 
Each sequence contains $T=9$ steps.

Evaluating the quality of an interpolation is straightforward for simple artificial objects, 
e.g. 2D lines whose length and orientation can be easily measured~\cite{ae_interp_ACAI_2019}.
However, it is harder to define what a "good" audio interpolation is.
Thus, this work relies on Timbre Toolbox~\cite{timbretoolbox_jasa11} to extract audio features engineered by experts.
Timbre features (related to envelope, spectrum, ...) have been extracted for all rendered audio files.
All available features but \textit{Noisiness}, which was almost constant to an inconsistent 1.0 value, have been included in the results (Fig.~\ref{fig:main_detailed_interp_results}).
A logarithm function has been applied to spectral features in Hz, for values to lie on an approximately linear perceptual scale.
Timbre Toolbox features are computed inside several time frames (slices), such that multiple values are available for each feature of a given audio file.
Following Peeters \textit{et al.}~\cite{timbretoolbox_jasa11}, only the median value and Inter-Quartile Range (IQR) are used.

Similar to~\cite{ae_interp_ACAI_2019, transformer_vae_3Dmotion_ICCV21}, two metrics have been computed for each interpolation sequence: smoothness and non-linearity.
The smoothness of a feature is defined as the RMS of the second-order derivative of this feature's values across the sequence. 
Non-linearity is the RMS distance between measured feature values, and an ideal linear regression from start to end values of the feature.

\subsection{Results}

The most common preset interpolation method, which is implemented in some commercial synthesizers, 
consists in performing an independent linear interpolation for each synthesis parameter.
This can lead to smooth interpolations, because synthesis controls are usually mapped to a perceptual scale, e.g. a log-scale for frequencies and amplitudes. Thus, this method has been considered as the reference interpolation. 

Fig.~\ref{fig:main_detailed_interp_results} shows the smoothness of audio features, for interpolations computed by our model and using the reference method.
Thirty-five features are significantly smoother (one-sided Wilcoxon signed-rank test, p-value $< 0.05$), 
and the average smoothness  is improved, i.e. reduced, by $12.6$\% (Table~\ref{tab:results}). 
This table also presents improved nonlinearity results, not displayed in Fig.~\ref{fig:main_detailed_interp_results} due to space constraints.
The companion website\footnote{https://gwendal-lv.github.io/spinvae/} presents examples of interpolations between presets,
and also extrapolations beyond test presets.

\begin{table}[!ht]
\centering
\caption{Performance of different interpolation models compared to the reference linear per-parameter interpolation.
	Number of significantly improved features (out of $46$ total)
	and average feature value variation (lower is better) for the Smoothness and Nonlinearity metrics.
}
\vspace{2mm}
\begin{tabular}{lcccc@{\hspace{0mm}}}

	  & \multicolumn{2}{c}{\textbf{\# improved} } & \multicolumn{2}{c}{\textbf{Average}} \\
	\textbf{Model} &  \multicolumn{2}{c}{\textbf{features}} & \multicolumn{2}{c}{\textbf{variation (\%)}} \\
	 & \textbf{Smooth.} & \textbf{Nonlin.} & \textbf{Smooth.} & \textbf{Nonlin.}  \\
	\hline
	\textbf{SPINVAE} & $\mathbf{35}$ & $38$ & $\mathbf{-12.6}$ & $-12.3$ \\
	Preset-only & $25$ & $30$ & $-4.6$ & $-6.4$  \\
	Sound match. & $8$ & $7$ & $+66.8$ & $+29.7$  \\
	\hline
	DLM 2 & $31$ & $37$ & $-8.2$ & $-10.4$   \\
	DLM 4 & $30$ & $\mathbf{40}$ & $-9.2$ & $-14.5$  \\
	Softmax & $23$ & $\mathbf{40}$ & $-1.2$  & $\mathbf{-15.6}$ \\
	\hline
	MLP  & $18$ & $27$ & $+21.0$ & $-1.8$  \\
    LSTM & $15$ & $1$ & $+123$ & $+93.5$  \\
	\hline
\end{tabular}
\label{tab:results}
\end{table}

\section{Ablation Study}\label{sec:ablation_study}

Table~\ref{tab:results} demonstrates that our interpolation model outperforms other related architectures, 
and provides insight into its ability to increase performance.

Models from the first section of Table~\ref{tab:results} are general architecture variants of SPINVAE.
The Preset-only model does not auto-encode $\x$, i.e. it is a Transformer-VAE~\cite{transformer_vae_3Dmotion_ICCV21, transformer_vae_music_ICASSP20} applied to presets.
Compared to the reference, it improves the interpolation but does not perform as well as the bimodal SPINVAE.
This indicates that learning audio representations alongside preset representations is well suited to this interpolation task.
The Sound matching model uses the same architecture as~\cite{esling2020flowsynth, presetgenVAE_2021}, 
which can be obtained by setting the preset encoder outputs to zero i.e. $q(\z | \x, \u) = q(\z | \x)$. 
However, we could not use the exact same decoder model as these previous works, because they rely on bijective networks which impose strong constraints on the latent dimension and are expected to model continuous numerical synthesis parameters only. 
The degraded performance of this Sound matching model probably comes from a discrepancy between the learned audio-only latent representations, and the corresponding decoded presets.

The second section of Table~\ref{tab:results} presents results obtained with different probability distributions used to model numerical synthesis parameters.
DLM 2 and 4 designate DLMs of two and four components, respectively (instead of three for SPINVAE), while Softmax indicates that discrete numerical values are learned as categories. 
The latter had been used to improve sound matching performance~\cite{sound2synth_2022, presetgenVAE_2021, inversynth2019}.
The interpolation performance is slightly reduced when using DLM 2 distributions, which are not flexible enough to model the data. 
DLM 4 and Softmax, nonetheless, improve linearity while degrading smoothness. 
However, subjective listening tests seemed to indicate that smoothness is more important to interpolation quality than linearity. 
Therefore, SPINVAE uses the DLM distribution, with three components rather than four.

Models from the last section of Table~\ref{tab:results} encode and decode presets using MLP or LSTM networks.
They perform poorly, which confirms that Transformer blocks are better suited to handle synthesizer presets.

\section{Conclusion}

The SPINVAE architecture has been introduced to auto-encode synthesizer presets and audio simultaneously, 
in order to perform interpolation between presets. 
Sequences of synthesized sounds, obtained from interpolated presets, were evaluated by computing the smoothness and nonlinearity of 46 audio features.
The evaluation demonstrated that SPINVAE outperforms related architectures, e.g. generative sound matching models.
It is also the first model to encode and decode presets using Transformer blocks, and to apply DLM distributions to presets.
An ablation study showed that these two elements helped improve the interpolation.

The proposed model was trained on a complex and non-differentiable FM synthesis process, and can be virtually applied to any commercial synthesizer.
It can be integrated into synthesizer plugins for live preset interpolation or sound design.
Combined with sound matching, the model could even perform interpolations between any re-synthesized recorded sounds.

\vfill\pagebreak

\bibliographystyle{IEEEbib}
\bibliography{icassp23}

\end{document}